\documentclass{IEEEconf}
\usepackage{amsthm}
\usepackage[pdftex,bookmarks=true]{hyperref}
\newtheorem{prop}{Proposition}
\theoremstyle{definition}
\newtheorem{exmp}{Example}

\begin{document}

\title{Discussion on Supervisory Control by Solving Automata Equation}

\author{
Victor Bushkov \\
\begin{affiliation}
Department of EECS \\
Tomsk State University \\
Tomsk, 634050, Russia
\end{affiliation} \\
\email{v.bushkov@gmail.com}
\and
Nina Yevtushenko \\
\begin{affiliation}
Department of EECS \\
Tomsk State University \\
Tomsk, 634050, Russia
\end{affiliation} \\
\email{ninayevtushenko@yahoo.com}
\and
Tiziano Villa \\
\begin{affiliation}
Department of CS \\
University of Verona \\
37134 Verona, Italy
\end{affiliation} \\
\email{tiziano.villa@univr.it}
}

\maketitle

\begin{abstract}
In this paper we consider the supervisory control problem through language equation solving. The equation solving approach allows to deal with more general topologies and to find a largest supervisor which can be used as a reservoir for deriving an optimal controller. We introduce the notions of solutions under partial controllability and partial observability, and we show how supervisory control problems with partial controllability and partial observability can be solved by employing equation solving methods.
\end{abstract}

\section{Introduction}\label{sec:intro}

The problem of supervisory control is well known \cite{c1,c2}. A discrete event system $P$, called the plant, should be controlled by a supervisor $C$ in order to meet the specification $S$. In other words, we are required to construct a supervisor (also called a controller) $C$ that combined with $P$ satisfies $S$. In this paper, we assume that all the behaviors are described by regular languages and thus, can be represented by finite automata.

Sometimes more supervisor restrictions are imposed. When considering partial controllability some actions of the plant cannot be disabled by a supervisor, while under partial observability some plant actions cannot be observed by a supervisor.

According to the problem statement, the problem of constructing a supervisor is very close to the problem of solving a language (or an automata) equation and it is known \cite{c3} how to derive a largest solution to the automata equation $\textsf{\small{\textit{P}}} \diamond \textsf{\small{\textit{X}}} \cong \textsf{\small{\textit{S}}}$, where $\textsf{\small{\textit{S}}}$ is the behavior of the overall system, $\textsf{\small{\textit{P}}}$ is the behavior of the known part of the system, $\textsf{\small{\textit{X}}}$ is the unknown component, and $\cong$  is a parallel composition operator. However, these methods cannot be directly used to solve the supervisory control problem due to the presence of uncontrollable and unobservable events (which are usually defined in a different way for language and automata equations). In this paper, we describe particular solutions of an automata equation under such limitations.

\section{Preliminaries}\label{sec:prel}

An \emph{automaton} is a quintuple $\textsf{\small{\textit{P}}} = (P, \Sigma, p_{0}, T_{P}, F_{P})$, where $P$ is a finite non-empty set of states with the initial state $p_{0}$ and the subset $F_{P}$ of final (accepting) states, $\Sigma$ is an alphabet, and $T_{P} \subseteq P \times \Sigma \times P$ is a transition relation which is extended to words in a usual way. The language accepted by $\textsf{\small{\textit{P}}}$  is the set $L(\textsf{\small{\textit{P}}}) = \{(\alpha \in \Sigma^{\ast}: \exists p \in F_{P} (p_{0}, \alpha, p) \in T_{p})\}$. An automaton is \emph{trim} if from each state a final state can be reached. An automaton with a prefix-closed language is a prefix-closed automaton. Moreover, automaton $Init(\textsf{\small{\textit{P}}})$ is a trim automaton with the language that is the prefix-closure of the language of \textsf{\small{\textit{P}}}. An automaton \textsf{\small{\textit{R}}} is a \emph{reduction} of an automaton \textsf{\small{\textit{P}}} if $L(\textsf{\small{\textit{R}}}) \subseteq L(\textsf{\small{\textit{P}}})$ (written, $\textsf{\small{\textit{R}}} \leq \textsf{\small{\textit{P}}}$). If $L(\textsf{\small{\textit{R}}}) = L(\textsf{\small{\textit{P}}})$ then automata $\textsf{\small{\textit{R}}}$ and $\textsf{\small{\textit{P}}}$ are equivalent (written, $\textsf{\small{\textit{R}}} \cong \textsf{\small{\textit{P}}}$). Given two automata $\textsf{\small{\textit{P}}}$ and $\textsf{\small{\textit{C}}}$ with languages $L(\textsf{\small{\textit{P}}}) \subseteq \Sigma_{1}^{\ast}$ and $L(\textsf{\small{\textit{C}}}) \subseteq \Sigma_{1}^{\ast}$, let $E$ be a non-empty subset of $\Sigma_{1} \cup \Sigma_{2}$. The \emph{parallel composition} $\textsf{\small{\textit{P}}} \diamond_{E} \textsf{\small{\textit{C}}}$ is the automaton $(\textsf{\small{\textit{P}}}_{\Uparrow\Sigma_{2}} \cap \textsf{\small{\textit{C}}}_{\Uparrow\Sigma_{1}})_{\Downarrow E}$. When clear from the context, instead of $\textsf{\small{\textit{P}}} \diamond_{E} \textsf{\small{\textit{C}}}$ we simply write $\textsf{\small{\textit{P}}} \diamond  \textsf{\small{\textit{C}}}$. If $E = \Sigma_{1} = \Sigma_{2}$, then $\textsf{\small{\textit{P}}} \diamond_{E} \textsf{\small{\textit{C}}} \cong \textsf{\small{\textit{P}}} \cap \textsf{\small{\textit{C}}}$ with the language $L(\textsf{\small{\textit{P}}}) \cap L(\textsf{\small{\textit{C}}})$. Correspondingly, given the automaton S with the language $L(\textsf{\small{\textit{S}}}) \subseteq E^{\ast}$ we consider an automata equation $\textsf{\small{\textit{P}}} \diamond_{E} \textsf{\small{\textit{X}}} \cong \textsf{\small{\textit{S}}}$, where $\textsf{\small{\textit{X}}}$ is an unknown automaton with the language over alphabet $\Sigma_{2}$. An automaton $C$ with the language over alphabet $\Sigma_{2}$ is a \emph{solution} to the equation if $\textsf{\small{\textit{P}}} \diamond_{E} \textsf{\small{\textit{C}}} \cong \textsf{\small{\textit{S}}}$. It is known that a solvable equation $\textsf{\small{\textit{P}}} \diamond_{E} \textsf{\small{\textit{X}}} \cong \textsf{\small{\textit{S}}}$ has a largest solution $\textsf{\small{\textit{M}}} = \overline{\textsf{\small{\textit{P}}} \diamond_{E} \overline{\textsf{\small{\textit{S}}}}}$\cite{c3}: the language of each solution is contained in the language of a largest solution. As usual, a number of particular solutions can be considered when solving automata equation \cite{c3}. In this paper, all automata in an automata equation are assumed to be trim.

\section{Supervisor synthesis by solving automata equations}\label{sec:dutomata equations and supervisory control problems}

\subsection{Describing the set of supervisors}\label{subsec:describing supervisors}

Let $\textsf{\small{\textit{P}}} = (P, \Sigma, p_{0}, T_{P}, F_{P})$ and $\textsf{\small{\textit{S}}} = (S, \Sigma, s_{0}, T_{S}, F_{S})$ be trim automata which describe the plant and the specification behavior, correspondingly. The problem is to derive a supervisor $\textsf{\small{\textit{C}}} = (C, \Sigma, c_{0}, T_{C}, F_{C})$ with a prefix-closed language such that $\textsf{\small{\textit{P}}} \diamond \textsf{\small{\textit{C}}} \cong \textsf{\small{\textit{S}}}$. Since $\textsf{\small{\textit{P}}}$, $\textsf{\small{\textit{S}}}$ and $\textsf{\small{\textit{C}}}$ are defined over the same alphabet, we are required to solve the equation $\textsf{\small{\textit{P}}} \cap \textsf{\small{\textit{X}}} \cong \textsf{\small{\textit{S}}}$. Then the equation is known to have a largest solution $\overline{\textsf{\small{\textit{P}}} \cap \overline{\textsf{\small{\textit{S}}}}} \cong \overline{\textsf{\small{\textit{P}}}} \cup \textsf{\small{\textit{S}}}$ and we denote by $(\overline{\textsf{\small{\textit{P}}}} \cup \textsf{\small{\textit{S}}})^{pref}$ the largest subautomaton of $ \overline{\textsf{\small{\textit{P}}}} \cup \textsf{\small{\textit{S}}}$ with a prefix-closed language. Thus, there exists a supervisor $\textsf{\small{\textit{C}}}$ such that $\textsf{\small{\textit{P}}} \cap \textsf{\small{\textit{C}}} \cong \textsf{\small{\textit{S}}}$ iff $\textsf{\small{\textit{P}}} \cap (\overline{\textsf{\small{\textit{P}}}} \cup \textsf{\small{\textit{S}}})^{pref} \cong \textsf{\small{\textit{S}}}$. On the other hand, for each $\textsf{\small{\textit{C}}}$ such that $\textsf{\small{\textit{P}}} \cap \textsf{\small{\textit{C}}} \cong \textsf{\small{\textit{S}}}$ it holds that $L($\textsf{\small{\textit{C}}}$) \supseteq L($\textsf{\small{\textit{S}}}$)$ and thus, the following statement holds.

\begin{prop}
Given the plant $\textsf{\small{\textit{P}}}$ and the specification $\textsf{\small{\textit{S}}}$, there exists a supervisor $\textsf{\small{\textit{C}}}$ such that $\textsf{\small{\textit{P}}} \cap \textsf{\small{\textit{C}}} \cong \textsf{\small{\textit{S}}}$ iff $\textsf{\small{\textit{P}}} \cap (\overline{\textsf{\small{\textit{P}}}} \cup \textsf{\small{\textit{S}}})^{pref} \cong \textsf{\small{\textit{S}}}$. Moreover, when a supervisor exists an automaton $\textsf{\small{\textit{C}}}$ with a prefix-closed language is a supervisor iff $Init(\textsf{\small{\textit{S}}}) \leq \textsf{\small{\textit{C}}} \leq (\overline{\textsf{\small{\textit{P}}}} \cup \textsf{\small{\textit{S}}})^{pref}$.
\end{prop}

However, not every supervisor is of practical use. If the languages of the plant and the specification are not prefix-closed then the intersection $\textsf{\small{\textit{P}}} \cap \textsf{\small{\textit{C}}}$ is not necessary a trim automaton and thus, a deadlock or a livelock can occur during the joint work of the plant and the supervisor. To escape such drawbacks the notion of a progressive (non-blocking) supervisor is used. A supervisor $\textsf{\small{\textit{C}}}$ is progressive if the automaton $\textsf{\small{\textit{P}}} \cap \textsf{\small{\textit{C}}}$ is trim. If the equation $\textsf{\small{\textit{P}}} \cap \textsf{\small{\textit{X}}} \cong \textsf{\small{\textit{S}}}$ is solvable then a supervisor with language $Init(L(\textsf{\small{\textit{S}}}))$ is progressive. However, it is not always the case for the supervisor $(\overline{\textsf{\small{\textit{P}}}} \cup \textsf{\small{\textit{S}}})^{pref}$.

\begin{exmp}
Consider $\textsf{\small{\textit{P}}}$ and $\textsf{\small{\textit{S}}}$ with the languages $\{a, abc\}$ and $\{a\}$ defined over the alphabet $\{a, b, c\}$, correspondingly. The language of a largest supervisor $\textsf{\small{\textit{C}}}$ has each word except of $abc$ and all continuations of this word; however, $\textsf{\small{\textit{C}}}$ is not progressive, since the automaton $\textsf{\small{\textit{P}}} \cap \textsf{\small{\textit{C}}}$ is not trim.
\end{exmp}

The notion of a progressive supervisor coincides with the notion of a progressive solution of an automata equation \cite{c4} and thus, a largest progressive supervisor exists if the equation $\textsf{\small{\textit{P}}} \cap \textsf{\small{\textit{X}}} \cong \textsf{\small{\textit{S}}}$ is solvable. A largest progressive supervisor can be derived in the same way as a largest progressive solution is derived, i.e., by deleting 'bad' sequences from the language of the automaton $(\overline{\textsf{\small{\textit{P}}}} \cup \textsf{\small{\textit{S}}})^{pref}$. A sequence is 'bad' if it is in the language $Init(L(\textsf{\small{\textit{P}}}))$ while having no continuation in $L(\textsf{\small{\textit{S}}})$. For this reason, differently from the general case of the largest progressive solution to automata equations the following proposition holds.

\begin{prop}
Each automaton $\textsf{\small{\textit{C}}}$ with a prefix-closed language is a progressive supervisor iff $Init(\textsf{\small{\textit{S}}}) \leq \textsf{\small{\textit{C}}} \leq (\overline{Init(\textsf{\small{\textit{P}}})} \cup Init(\textsf{\small{\textit{S}}}))^{pref}$, where $(\overline{Init(\textsf{\small{\textit{P}}})} \cup Init(\textsf{\small{\textit{S}}}))^{pref}$ is the largest progressive supervisor.
\end{prop}

\subsection{Describing the set of supervisors under partial controllability}\label{subsec:describing supervisors partial controll}

When talking about partial controllability one assumes that a supervisor cannot prevent the occurrence of \emph{uncontrollable} actions, i.e., alphabet $\Sigma$ is partitioned into two subsets $\Sigma_{c}$ and $\Sigma_{uc}$, where $\Sigma_{c}$ and $\Sigma_{uc}$ are the sets of controllable and uncontrollable actions, respectively. Given an automaton $\textsf{\small{\textit{C}}}$ over alphabet $\Sigma$, we obtain the $\Sigma_{uc}$-extension $\textsf{\small{\textit{C}}}^{\Uparrow\Sigma_{uc}}$ of $\textsf{\small{\textit{C}}}$ by adding at each state of $\textsf{\small{\textit{C}}}$ a self-loop labeled with each action $a \in \Sigma_{uc}$ such that there is no transition from this state under action $a$.

A solution $\textsf{\small{\textit{C}}}$ of the equation $\textsf{\small{\textit{P}}} \cap \textsf{\small{\textit{X}}} \cong \textsf{\small{\textit{S}}}$ is a \emph{solution under partial controllability} if $\textsf{\small{\textit{C}}}^{\Uparrow\Sigma_{uc}}$ is a solution of the equation $\textsf{\small{\textit{P}}} \cap \textsf{\small{\textit{X}}} \cong \textsf{\small{\textit{S}}}$. The following statement establishes necessary and sufficient conditions for the equation solvability under partial controllability.

\begin{prop}
Given solvable equation $\textsf{\small{\textit{P}}} \cap \textsf{\small{\textit{C}}} \cong \textsf{\small{\textit{S}}}$.
\begin{itemize}
\item[(i)] The equation is solvable under partial controllability iff $Init(L(\textsf{\small{\textit{S}}}))(\Sigma_{uc})^{\ast} \subseteq L(\overline{\textsf{\small{\textit{P}}}}) \cup L(\textsf{\small{\textit{S}}})$.
\item[(ii)] If the equation $\textsf{\small{\textit{P}}} \cap \textsf{\small{\textit{X}}} \cong \textsf{\small{\textit{S}}}$ is solvable under partial controllability, then it has a largest solution under partial controllability.
\end{itemize}
\end{prop}

However, it may occur that neither $Init(\textsf{\small{\textit{S}}})$ nor $(\overline{\textsf{\small{\textit{P}}}} \cup \textsf{\small{\textit{S}}})^{pref}$ are solutions under partial controllability.

\begin{exmp}
Consider $\textsf{\small{\textit{P}}}$ and $\textsf{\small{\textit{S}}}$ with the languages $\{\epsilon, ba\}$ and $\{\epsilon\}$ over $\Sigma = \{a, b\}$, correspondingly. Let  $\Sigma_{uc} = \{a\}$. The language of $(\overline{\textsf{\small{\textit{P}}}} \cup \textsf{\small{\textit{S}}})^{pref}$ contains all words over $\Sigma$, except those that have $ba$ as a prefix. Then the language of $((\overline{\textsf{\small{\textit{P}}}} \cup \textsf{\small{\textit{S}}})^{pref})^{\Uparrow\Sigma_{uc}}$ contains the word $ba$. As a result, $((\overline{\textsf{\small{\textit{P}}}} \cup \textsf{\small{\textit{S}}})^{pref})^{\Uparrow\Sigma_{uc}}$ is not a solution of the equation $\textsf{\small{\textit{P}}} \cap \textsf{\small{\textit{X}}} \cong \textsf{\small{\textit{S}}}$.
\end{exmp}

\begin{exmp}
Consider $\textsf{\small{\textit{P}}}$ and $\textsf{\small{\textit{S}}}$ with the languages $\{\epsilon, b, ab\}$ and $\{\epsilon\}$ over $\Sigma = \{a, b\}$, correspondingly. Let  $\Sigma_{uc} = \{a\}$. The automaton $Init(\textsf{\small{\textit{S}}})^{\Uparrow\Sigma_{uc}}$ is not a solution, since its language contains the word $ab$. But the equation $\textsf{\small{\textit{P}}} \cap \textsf{\small{\textit{X}}} \cong \textsf{\small{\textit{S}}}$ is solvable under partial controllability, for example, an automaton with the language $\{ \epsilon, b, a\}$ is a solution under partial controllability.
\end{exmp}

 A largest solution under partial controllability can be obtained by iteratively eliminating each state $st$ of the automaton $(\overline{\textsf{\small{\textit{P}}}} \cup \textsf{\small{\textit{S}}})^{pref}$, such that from $st$ there are no transitions under some uncontrollable action, until every state has a transition for every uncontrollable action; if the resulting automaton is not a solution, then the equation has no solutions and the intersection of the resulting automaton with the plant gives the largest controllable behavior we could achieve. However, as the following proposition states, if the languages of $\textsf{\small{\textit{P}}}$ and $\textsf{\small{\textit{S}}}$ are prefix-closed, then there is no need for trimming of the automaton $(\overline{\textsf{\small{\textit{P}}}} \cup \textsf{\small{\textit{S}}})^{pref}$.

\begin{prop}
If the languages of $\textsf{\small{\textit{P}}}$ and $\textsf{\small{\textit{S}}}$ are prefix-closed and the equation $\textsf{\small{\textit{P}}} \cap \textsf{\small{\textit{X}}} \cong \textsf{\small{\textit{S}}}$ is solvable under partial controllability then an automaton $\textsf{\small{\textit{C}}}$ with a prefix-closed language is a supervisor iff $Init(\textsf{\small{\textit{S}}}) \leq \textsf{\small{\textit{C}}} \leq (\overline{\textsf{\small{\textit{P}}}} \cup \textsf{\small{\textit{S}}})^{pref}$.
\end{prop}

A solution $\textsf{\small{\textit{C}}}$ of the equation $\textsf{\small{\textit{P}}} \cap \textsf{\small{\textit{X}}} \cong \textsf{\small{\textit{S}}}$ is a \emph{progressive solution under partial controllability} if $\textsf{\small{\textit{C}}}^{\Uparrow\Sigma_{uc}}$ is a progressive solution of the equation $\textsf{\small{\textit{P}}} \cap \textsf{\small{\textit{X}}} \cong \textsf{\small{\textit{S}}}$. Unlike the case when all events are controllable, a progressive solution of the equation is not always progressive under partial controllability.

\begin{exmp}
Let $\Sigma_{uc} = \{a\}$, $L(\textsf{\small{\textit{P}}}) = \{ \epsilon, ab\}$, $L(\textsf{\small{\textit{S}}}) = \{ \epsilon\}$. Then automaton $\textsf{\small{\textit{C}}}$ with the language $L(\textsf{\small{\textit{C}}}) = \{ \epsilon\}$ is a progressive solution of the equation and is a solution under partial controllability; however, $\textsf{\small{\textit{C}}}$ is not a progressive solution under partial controllability.
\end{exmp}

Nevertheless, it turns out that if the equation $\textsf{\small{\textit{P}}} \cap \textsf{\small{\textit{X}}} \cong \textsf{\small{\textit{S}}}$ has a progressive solution, then a progressive solution under partial controllability is equivalent to a corresponding progressive solution.

\begin{prop}
If the equation $\textsf{\small{\textit{P}}} \cap \textsf{\small{\textit{X}}} \cong \textsf{\small{\textit{S}}}$ has a progressive solution under partial controllability and $\textsf{\small{\textit{C}}}$ is a prefix-closed solution then:
\begin{itemize}
\item[(i)] $\textsf{\small{\textit{C}}}$ is a progressive solution under partial controllability iff $\textsf{\small{\textit{C}}}$ is a progressive solution.
\item[(ii)] $\textsf{\small{\textit{C}}}$ is a progressive solution under partial controllability iff $Init(\textsf{\small{\textit{S}}}) \leq \textsf{\small{\textit{C}}} \leq (\overline{\textsf{\small{\textit{P}}}} \cup \textsf{\small{\textit{S}}})^{pref}$.
\end{itemize}
\end{prop}

\subsection{Describing the set of supervisors under partial observability}\label{subsec:describing supervisors partial observ}

When talking about partial observability one assumes that the supervisor cannot `see' the occurrence of unobservable actions, i.e., the $\Sigma$ is partitioned into two subsets $\Sigma_{o}$ and $\Sigma_{uo}$, where $\Sigma_{o}$ and $\Sigma_{uo}$ are the sets of observable and unobservable actions, respectively. However, the plant can observe each action of the supervisor and correspondingly under complete controllability the plant can execute an action iff both, the plant and the supervisor, are ready to execute the action at their current states. After executing an action unobservable by a supervisor the plant moves to the next state while the supervisor remains at its current state. If an action is observable by a supervisor then both, the plant and the supervisor, execute a corresponding transition. Here we notice that in general case, partial controllability and observability are considered independently. Uncontrollable actions can be observable while controllable actions can be unobservable and vice versa. Since a supervisor cannot `see' unobservable actions, it is necessary to impose additional conditions in order to have a solution of the equation $\textsf{\small{\textit{P}}} \cap \textsf{\small{\textit{X}}} \cong \textsf{\small{\textit{S}}}$ under partial observability.

 Given an automaton $\textsf{\small{\textit{C}}}$ over alphabet $\Sigma$, we obtain the $\Sigma_{uo}$-folding $\textsf{\small{\textit{C}}}^{\Downarrow\Sigma_{uo}}$ of $\textsf{\small{\textit{C}}}$ by replacing each transition $(c_{1}, a, c_{2})$ of $\textsf{\small{\textit{C}}}$, such that $a \in \Sigma_{uo}$, with a self-loop at state $c_{1}$.

Let $\Sigma = \Sigma_{o} \cup \Sigma_{uo}$. A solution $\textsf{\small{\textit{C}}}$ of the equation $\textsf{\small{\textit{P}}} \cap \textsf{\small{\textit{X}}} \cong \textsf{\small{\textit{S}}}$ is a \emph{solution under partial observability} if $\textsf{\small{\textit{C}}}^{\Downarrow\Sigma_{uo}}$ is a solution of the equation $\textsf{\small{\textit{P}}} \cap \textsf{\small{\textit{X}}} \cong \textsf{\small{\textit{S}}}$.

Given an automaton $\textsf{\small{\textit{C}}}$ over alphabet $\Sigma = \Sigma_{o} \cup \Sigma_{uo}$, we obtain the automaton $\textsf{\small{\textit{C}}}_{real}$ by adding a self-loop at each state $\{c_{1},\ldots, c_{n}\}$ of the deterministic restriction $\textsf{\small{\textit{C}}}_{\Downarrow\Sigma_{o}}$ labeled with each action $a \in \Sigma_{uo}$ such that from some state $c_{i} \in \{c_{1},\ldots, c_{n}\}$ there is a transition under $a$ in the automaton $\textsf{\small{\textit{C}}}$.

\begin{prop}
The equation $\textsf{\small{\textit{P}}} \cap \textsf{\small{\textit{C}}} \cong \textsf{\small{\textit{S}}}$ is solvable under partial observability iff $(Init(L(\textsf{\small{\textit{S}}})))_{real} \subseteq L(\overline{\textsf{\small{\textit{P}}}}) \cup L(\textsf{\small{\textit{S}}})$.
\end{prop}

Unfortunately, the union of two solutions under partial observability is not necessary a solution under partial observability and thus, a largest solution does not exist under partial observability. We demonstrate this by a simple example.

\begin{exmp}
Let $\Sigma_{o} = \{b\}$, $L(\textsf{\small{\textit{P}}}) = \{ \epsilon, ab\}$, and $L(\textsf{\small{\textit{S}}}) = \{ \epsilon\}$. Consider automata $\textsf{\small{\textit{C}}}_{1}$ and $\textsf{\small{\textit{C}}}_{2}$ with the languages $L(\textsf{\small{\textit{C}}}_{1}) = \{ \epsilon, a\}$ and $L(\textsf{\small{\textit{C}}}_{2}) = \{ \epsilon, b\}$ which are solutions of the equation $\textsf{\small{\textit{P}}} \cap \textsf{\small{\textit{X}}} \cong \textsf{\small{\textit{S}}}$. The automaton $\textsf{\small{\textit{C}}}_{1} \cup \textsf{\small{\textit{C}}}_{2}$ has the language $\{\epsilon , a, b\}$ and thus, the language of $(\textsf{\small{\textit{C}}}_{1} \cup \textsf{\small{\textit{C}}}_{2})^{\Downarrow\Sigma_{uo}}$ equals ${\epsilon , a^{\ast}, a^{\ast}b}$. The intersection of this language with $L(\textsf{\small{\textit{P}}})$ has the word $ab$ which is not contained in $L(\textsf{\small{\textit{S}}})$, i.e., $\textsf{\small{\textit{C}}}_{1} \cup \textsf{\small{\textit{C}}}_{2}$ is not a supervisor under partial observability.
\end{exmp}

A solution $\textsf{\small{\textit{C}}}$ of the equation $\textsf{\small{\textit{P}}} \cap \textsf{\small{\textit{X}}} \cong \textsf{\small{\textit{S}}}$ is a \emph{progressive solution under partial observability} if $\textsf{\small{\textit{C}}}^{\Downarrow\Sigma_{uo}}$ is a progressive solution of the equation $\textsf{\small{\textit{P}}} \cap \textsf{\small{\textit{X}}} \cong \textsf{\small{\textit{S}}}$. A solution under partial observability that is a progressive solution of the equation is not necessary a progressive solution under partial observability, even when the equation has progressive solutions under partial observability. Moreover, a progressive solution under partial observability is not always a progressive solution of the equation.

A solution $\textsf{\small{\textit{C}}}$ of the equation $\textsf{\small{\textit{P}}} \cap \textsf{\small{\textit{X}}} \cong \textsf{\small{\textit{S}}}$ is a \emph{solution under partial controllability and observability} if $(\textsf{\small{\textit{C}}}^{\Downarrow\Sigma_{uo}})^{\Uparrow\Sigma_{uc}}$ is a solution of the equation. A solution $\textsf{\small{\textit{C}}}$ of the equation $\textsf{\small{\textit{P}}} \cap \textsf{\small{\textit{X}}} \cong \textsf{\small{\textit{S}}}$ is a \emph{progressive solution under partial controllability and observability} if $(\textsf{\small{\textit{C}}}^{\Downarrow\Sigma_{uo}})^{\Uparrow\Sigma_{uc}}$ is a progressive solution of the equation. It can be shown that $(\textsf{\small{\textit{C}}}^{\Downarrow\Sigma_{uo}})^{\Uparrow\Sigma_{uc}} \cong (\textsf{\small{\textit{C}}}^{\Uparrow\Sigma_{uc}})^{\Downarrow\Sigma_{uo}}$. Sometimes a special case of partial controllability and observability is considered when each unobservable action cannot be controlled, i.e., $\Sigma_{uo} \subseteq \Sigma_{uc}$. In this case, there exists a largest supervisor.

\begin{exmp}
Let $\Sigma_{uo} = \Sigma_{uc} = \{b\}$, $L(\textsf{\small{\textit{P}}}) = \{ b, baa\}$, and $L(\textsf{\small{\textit{S}}}) = \{ b\}$. Then automaton $\textsf{\small{\textit{C}}}$ with the language $L(\textsf{\small{\textit{C}}}) = \{ \epsilon, b, ba\}$ is not a progressive solution of the equation, while $(\textsf{\small{\textit{C}}}^{\Downarrow\Sigma_{uo}})^{\Uparrow\Sigma_{uc}}$ is a progressive solution of the equation. Therefore $\textsf{\small{\textit{C}}}$ is a progressive solution under partial controllability and observability in spite of the fact that it is not progressive without the partial controllability and observability limitation.
\end{exmp}

\begin{exmp}
Let $\Sigma_{uo} = \Sigma_{uc} = \{b\}$, $L(\textsf{\small{\textit{P}}}) = \{ b, baa\}$, and $L(\textsf{\small{\textit{S}}}) = \{ b\}$. Then automaton $\textsf{\small{\textit{C}}}$ with the language $L(\textsf{\small{\textit{C}}}) = \{ \epsilon, b, a\}$ is a progressive solution of the equation, however, $(\textsf{\small{\textit{C}}}^{\Downarrow\Sigma_{uo}})^{\Uparrow\Sigma_{uc}}$ is not a progressive solution.
\end{exmp}

\begin{prop}
Let $\Sigma_{uo} \subseteq \Sigma_{uc}$ and let $\textsf{\small{\textit{Z}}}$ be automaton with the language $L(Init(\textsf{\small{\textit{S}}}))(\Sigma_{uc})^{\ast}$. The equation $\textsf{\small{\textit{P}}} \cap \textsf{\small{\textit{X}}} \cong \textsf{\small{\textit{S}}}$ is solvable under partial controllability and observability iff $L(\textsf{\small{\textit{Z}}}_{real}) \subseteq L(\overline{\textsf{\small{\textit{P}}}}) \cup L(\textsf{\small{\textit{S}}})$.
\end{prop}

\begin{prop}
If $\Sigma_{uo} \subseteq \Sigma_{uc}$ and the equation $\textsf{\small{\textit{P}}} \cap \textsf{\small{\textit{X}}} \cong \textsf{\small{\textit{S}}}$ is solvable under partial controllability and observability then there exists a largest solution under partial controllability and observability
\end{prop}

However, similar to the partial controllability the automaton $(\overline{\textsf{\small{\textit{P}}}} \cup \textsf{\small{\textit{S}}})^{pref}$ is not always a largest solution under partial controllability and observability, and in order to get a largest supervisor we need to trim $(\overline{\textsf{\small{\textit{P}}}} \cup \textsf{\small{\textit{S}}})^{pref}$.

\section{Conclusion}\label{sec:con}

In this paper, we have considered the problem of synthesizing a supervisor through automata equation solving. We have discussed progressive (non-blocking) supervisors as well as supervisors under partial controllability and observability and have shown that most special kinds of supervisors can be derived as proper solutions of a corresponding automata equation. Moreover, the complexity of solving a corresponding automata equation is not exponential as in general case but rather polynomial w.r.t. to the number of states of the plant and the specification. A largest proper supervisor (if exists) can be derived by trimming a largest solution to the automata equation. Moreover, differently from the general case each reduction of such trim automaton is also a supervisor. Each largest supervisor can be used as a reservoir for deriving an optimal supervisor that can be simpler than a traditional supervisor. Also, since the approach based on language equation solving can deal with more general topologies, this approach can be used for deriving supervisors when the plant, the specification and the supervisor have different sets of actions \cite{c5, c6}.

\section*{Acknowledgments}

The first author gratefully acknowledges support from the Bortnik Fund (contract 6360 ð/8858). The second author gratefully acknowledges support of RFBR-NSC (grant 06-08-89500).

\end{document}